\documentclass{article}
\usepackage{natbib}
\usepackage{amsthm,amsfonts,color}
\usepackage{mathptmx}
\usepackage{bm}
\usepackage{amsmath}
\usepackage[T1]{fontenc}
\usepackage{epsfig}
\usepackage{graphicx}
\usepackage{graphics}
\usepackage{lscape}
\usepackage{pdflscape}
\usepackage{amssymb,srcltx}

\newcommand{\x}{\mathbf{x}}

\newcommand{\Y}{\mathbf{Y}}

\newcommand{\V}{\mathbf{V}}
\newcommand{\z}{\mathbf{z}}
\newcommand{\Z}{\mathbf{Z}}

\newcommand{\aee}{\mathbf{a}}
\newcommand{\X}{\mathbf{X}}
\newcommand{\R}{\mathbf{R}}

\newcommand{\C}{\mathbf{C}}

\newcommand{\de}{\mathbf{d}}

\newcommand{\p}{\mathbf{p}}

\newcommand{\B}{\mathbf{B}}

\newcommand{\y}{\mathbf{y}}

\newcommand{\we}{\mathbf{w}}
\newcommand{\balpha}{\mbox{${\bm \alpha}$}}
\newcommand{\bmu}{\mbox{${\bm \mu}$}}

\newcommand{\bnu}{\mbox{${\bm \nu}$}}
\newcommand{\bSigma}{\mbox{${\bm \Sigma}$}}

\newcommand{\bepsilon}{\mbox{${\bm \epsilon}$}}
\newcommand{\bLambda}{\mbox{${\bm \Lambda}$}}
\newcommand{\bbeta}{\mbox{${\bm \beta}$}}
\newcommand{\btheta}{\mbox{${\bm \theta}$}}
\newcommand{\bTheta}{\mbox{${\bm \Theta}$}}
\newcommand{\bxi}{\mbox{${\bm \xi}$}}

\newcommand{\bDelta}{\mbox{${\bm \Delta}$}}
\newcommand{\bdelta}{\mbox{${\bm \delta}$}}

\newcommand{\be}{\mathbf{b}}

\newcommand{\blambda}{\mbox{${\bm \lambda}$}}
\newcommand{\bOmega}{\mbox{${\bm \Omega}$}}

\newcommand{\bomega}{\mbox{${\bm \omega}$}}

\newcommand{\bgamma}{\mbox{${\bm \gamma}$}}

\newtheorem{definition}{Definition}

\begin{document}
\title{Bayesian Measurement Error Models Using Finite
	Mixtures of Scale Mixtures of Skew-Normal Distributions}
\author{
\small Celso R\^omulo Barbosa Cabral
\thanks{Corresponding author.  Address for correspondence: Departamento de Estat\'{\i}stica,  Av. Gen. Rodrigo Oct\'avio, 6200, Coroado I. CEP 69080-900. Manaus, Amazonas, Brazil.  
e-mail adresses: \texttt{celsoromulo@ufam.edu.br} (C. R. B. Cabral),  \texttt{nelsonlima@ufam.edu.br} (Nelson Lima de Souza),   
\texttt{jeremias@ufam.edu.br } (Jeremias Le\~ao ) } 
\and \small  Nelson Lima de Souza \and \small Jeremias Le\~ao   \\
{ \em \small Departament of Statistics, Federal University of Amazonas, Brazil} \vspace*{0.1cm}
%  \author{The Corporation}
}
\date{}
\maketitle
\begin{abstract}
We present a proposal to deal with the non-normality issue in the context of regression models with measurement errors when both the response and the explanatory variable are observed with error. We extend the normal model by jointly modeling the unobserved covariate and the random errors by a finite mixture of scale mixture of skew-normal distributions. This approach allows us to model data with great flexibility, 
accommodating skewness, heavy tails, and multi-modality. %We develop a simple and efficient MCMC-type algorithm to perform Bayesian estimation of the parameters in the proposed model and compare the performance of our method with some competitors through the analysis of artificial and real data.
\vspace*{0.5cm}
\end{abstract}

\section{Introduction and Motivation}\label{sec motivation}
%\subsection{A Motivation Example}
{Let us  consider} the problem of modeling the relationship between two random variables $y$ and $x$ through a linear regression model, that is, 
$$
y= \alpha + \beta x,  
$$
{where} $\alpha$ and $\beta$ are parameters to be estimated. Supposing that these variables are unobservable, we assume that what we actually observe is
$$
X = x + \zeta, \,\,\, \mbox{and} \,\,\, Y= y + e, 
$$
where $\zeta$ and $e$ are random errors.  This is the so-called  \emph{measurement error (ME) model}. {There} is a vast literature regarding the inferential aspects of these kinds of models. Comprehensive reviews  can be found in \citet{fuller}, \cite{vaness} and \cite{Carrol.Ruppert.2006}. In general it is assumed that the variables $x$, $\zeta$ and $e$ are independent and normally distributed. However, there are situations when the true distribution of  the latent variable $x$ departs from normality; that is the case when skewness, {outliers} and multimodality are present. Then, the choice of more flexible models can be {a} useful alternative to the normal one in order to overcome possible drawbacks. To better understand the phenomena, consider the following  description of a real {dataset} (hereafter \emph{the SLE data}), which will be used to illustrate the applicability of the methods proposed in this article -- see Section \ref{sec_real_data}.

\begin{figure}[htbp] 
	\centering
%	\vspace{-0.6cm}
	\includegraphics[scale=0.50]{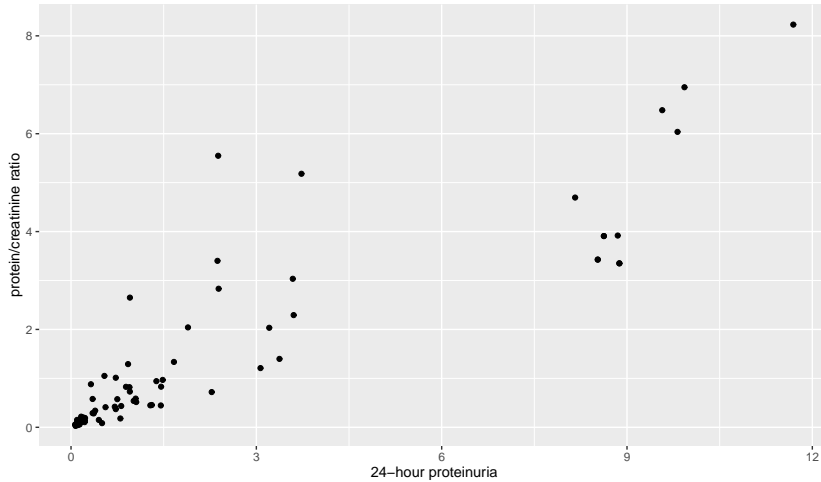}	
	\vspace{-0.3cm}
	\caption{24-hour proteinuria versus {protein}/creatinine ratio (both divided by 1000) for 75 SLE patients.}
	\label{lupus_dispersion}
\end{figure}

{Systemic lupus erythematosus} (SLE) is an autoimmune disease that affects many organs and systems. The prevalence and incidence of SLE vary with region, sex, age, ethnicity and time \citep{rees.2017}. Clinical manifestations involve skin and joint damages, inflammation of  membranes (pleura and pericardium), {as well as} neurological, hematological and renal alterations. Several studies show that SLE patients with renal disease {have} high mortality risk \citep{nieves.2016}. Thus, an important issue is to evaluate  the  renal function of SLE patients. In order to do so, a prospective study was performed by observing  patients with SLE  at the Rheumatology Service of the Ara\'ujo Lima {Outpatient Clinic} in Manaus, Brazil \citep{lima.2015}. The main goal was to  study the relationship between two tests, namely the protein/creatinine ratio taken from an isolated urine sample, and the 24-hour proteinuria. The protein/creatinine ratio test  is a simple test based on a sample from the first-morning urine. The 24-hour proteinuria test is considered a gold standard method, {as} an early and sensitive marker for the detection of possible renal damage. However, this latter method has  some disadvantages. For example, some patients can express annoyance {about the need} to collect {samples} for 24 hours. The two methods were applied {to} each of 75 patients {of} both genders, with {18} years old or more. Besides this, all the patients  fit the classification criteria for lupus defined by the American College of Rheumatology (ACR) and  the Systemic Lupus International Collaborating Clinics (SLICC). Suppose  that  $Y$ is the observed protein/creatinine ratio and $X$ is the observed 24-hour proteinuria. Figure~\ref{lupus_dispersion} shows a dispersion plot of $X$ {vs.} $Y$ (both divided by 1000), where {one} can clearly note departures from normality. In particular, one can see two distinct subgroups, due to a possible  unobserved heterogeneity.  In this case the distribution of the responses is possibly bimodal, and  the  usual normal regression model {cannot} be used. Our main goal in this work is to present a model with a flexible distribution for the latent covariate $x$ {so as to overcome} difficulties like these.    

 If the distribution of the latent variable $x$ departs from normality, the choice of more flexible  models can be {a} useful alternative to accommodate skewness, heavy tails or multimodality. Some works with this kind of approach are \cite{bolfArellano.1994},  \cite{galea.2005},  and \cite{Castro.Galea.2010}. All these authors consider ME models where the normal assumption for the joint distribution of $x$ and the measurement errors is replaced with the Student-$t$ distribution. The works of  \cite{Rocha.Loschi.Arellano.2016} and  \cite{Matos.Castro.Cabral.Lachos.2018}  consider an ME Student-$t$ model where the responses and predictors are both censored.

To allow more flexibility, \cite{lachos_etal:09b} and \cite{lachos_etal:10b} 
extended the ME model by considering that the joint distribution of $x$ and and the measurement errors  belongs to the class of the  \emph{scale mixtures of skew-normal distributions} (hereafter SMSN). The members of this family are extensions of classical symmetric  distributions. Thus, we have skew-normal, skew Student-$t$ and skew slash distributions, for example. The extension is obtained by the introduction of a shape parameter, as will {be seen} in Section \ref{subsec multivariate SN}. A related approach was also used by \cite{tomaya.castro.2018}, by supposing that the distribution of $x$ is skew Student-$t$ and the joint  distribution of the errors  is Student-$t$ in a heteroscedastic ME model.

When the source of {non-normality} is the presence 
of unobserved heterogeneity in the distribution of $x$, an alternative is to model this distribution by a finite mixture of normal densities, as in \cite{carroll_etal:99b}.
Also, as mentioned in \cite{mclachlan}, the great flexibility of finite mixtures of normal distributions allows {modeling} data when there is the simultaneous occurrence
of skewness, discrepant observations, and multimodality. But, even when using normal mixtures, one can have overestimation of the number of components (that is, the number of densities in the mixture) necessary to capture the asymmetric and/or heavy-tailed
nature of each subpopulation. To overcome this problem, \cite{cabral_etal:14}  assumed that  
the latent covariate and the random observational errors are jointly modeled by a finite mixture of skew Student-$t$ distributions, developing an EM-type algorithm {for} inference. Here, we extend this work, by supposing that this joint distribution is a finite mixture with components in the SMSN family. Also, a Bayesian inferential approach is adopted.

The remainder of the paper is organized as follows. In Section~\ref{subsec multivariate SN}, for the sake of completeness, we review the basic concepts of the skew-normal distribution and  its scale mixtures and also explore the concept of finite mixtures of distributions in this family. In Section~\ref{sec LMMSNMIX} we define our extension of the normal  measurement error  model. In Section~\ref{sec mcmc estimation} we develop the MCMC-type algorithm {for} Bayesian inference. In Section \ref{sec model selection} we discuss model selection and in Section \ref{sec aplication} {we give} numerical examples using both simulated and real data to illustrate the performance of the proposed {method}. 

\section{The  Skew Normal Distribution, the SMSN Family, and Mixtures}  \label{subsec multivariate SN}

The concept of skew normal (SN) distribution has a long history in the probability and statistical literature, going back to works like\,\,\cite{Birnbaum.1950} and \cite{Nelson.1964}. However, there is no doubt that the most popular approach {is that} given by Adelchi  Azzalini and colleagues when they presented extensions for the univariate and multivariate normal distributions in seminal papers like\, \citet{azzalini85} and \citet{azzalini96}, respectively, followed by   unification efforts in \cite{Arellano_Azzalini_2006} and \citet{Azzalini.Capitanio.SNBook}. Here we use a definition of skew normal distribution as a member of the \emph{fundamental skew-normal distribution family} (FUSN), presented in\,\cite{Arellano_Genton_2005}.

In what follows $\textrm{N}_q(\bmu,\bOmega)$ denotes the $q$-variate normal distribution with mean vector $\bmu$ e covariance matrix $\bSigma$,  $\textrm{N}_q(\cdot| \bmu,\bOmega)$ is the respective probability density function (pdf) and  $\Phi(\cdot)$ is the standard normal distribution function.   
\begin{definition} \label{def CFUSN}
	Let   $X_0 \sim \textrm{N}({0},1)$ and 
	$\X_1 \sim \textrm{N}_q(\bmu,\bSigma)$ be independent, where  $\bSigma$ is positive definite. Let $\bDelta$ be a $q$-dimensional vector. We say that the distribution of 
	$$
	\X =  \bDelta  |X_0| +   \X_1,
	$$	
	is skew normal with location vector $\bmu$, scale matrix $\bSigma$ and  shape vector  $\bDelta$. We use the notation $\X \sim \textrm{SN}_{q}(\bmu,\bSigma,\bDelta)$. 
\end{definition}
The distribution of  $\X$ belongs to the FUSN family because it is the   distribution of $\V|X_0>0$, where $\V=\bDelta  X_0 +   \X_1$. {Since} $\V$ and $X_0$ have {jointly} a  normal 
distribution, it is straightforward to prove that $\X$ has pdf  given by    
\begin{equation} \label{eqn SNpdf}
\textrm{SN}_q(\x|\bmu,\bSigma,\bDelta)= 2 \textrm{N}_q(\x| \bmu,\bOmega) \Phi(\blambda^\top(\y-\bmu)), 
\end{equation}
where 
\begin{equation} \label{eqn siglambda}
\bOmega = \bSigma+ \bDelta \bDelta^\top \, \, \,   \mbox{and}  \,\, \, \blambda = \frac{\bOmega^{-1} \bDelta}{(1 - \bDelta^\top \bOmega^{-1} \bDelta)^{1/2}}. 
\end{equation}
Also, we  recover $\bDelta$ and $\bSigma$ by  using 
\begin{equation*} \label{eqn parametrizacao beta v}
\bDelta=\bOmega\bdelta,  \,\,\,  \bSigma=\bOmega- \bOmega \bdelta \bdelta^{\top} \bOmega,
\end{equation*}
where $\bdelta=\blambda/(1+\blambda^{\top} \bOmega\blambda)^{1/2}$. Notice that  the case $\blambda= \mathbf{0}$ (equivalently $\bDelta=\mathbf{0}$) corresponds to the usual $q$-variate normal distribution. Also, the SN given in Definition \ref{def CFUSN} is the same used before in works like \cite{lachos_etal:09b} and \cite{cabral_etal:14}, defined by its pdf  as $f(\x
)= 2{\textrm{ N}_q(\x|\bmu,\bOmega)
	\Phi(\blambda^{*\top}\bOmega^{-1/2}(\x-\bmu))}$,
where $\bOmega^{-1/2}$ is the inverse of the square root of $\bOmega$. Equation \eqref{eqn SNpdf} is obtained through the  parameterization $\blambda=\bOmega^{-1/2} \blambda^{*}$.  

\begin{definition} \label{def SMSN}
	We say that the distribution of the $q$-dimensional random vector $\Y$ belongs to the family of \emph{scale mixtures of skew normal (SMSN) distributions} when
	\begin{equation} \label{eqn def SMSN}
	\Y= \bmu + U^{-1/2} \X,
	\end{equation}
	where $\bmu$ is a $q$-dimensional vector of constants,  $\X \sim
	\textrm{SN}_{q}(\mathbf{0},\bSigma,\bDelta)$ and  $U$ is a positive random variable independent of $\X$ having  distribution function
	$H(\cdot|\bnu)$. 
\end{definition}
Here $\bnu$ is a (possibly multivariate) parameter indexing the distribution of $U$, which is known as \emph{the scale factor}. $H(\cdot|\bnu)$ is {called}  the \emph{mixing distribution function}. We write $\Y \sim \textrm{SMSN}_q(\bmu,\bSigma,\bDelta,\bnu)$. By Definitions  \ref{def CFUSN} and  \eqref{def SMSN}, 
\begin{equation*} \label{eqn rep stoch SMSN first}
\Y|U=u\sim
\textrm{SN}_q(\bmu,u^{-1}\bSigma,u^{-1/2}\bDelta), 
\end{equation*}
which implies that the 
marginal pdf of $\Y$ is
\begin{equation*}\label{denSNI}
\textrm{SMSN}_{q}(\mathbf{y}|\bmu,\bSigma,\bDelta,\bnu)=
2\int^{\infty}_0{\textrm{N}_q(\mathbf{y}|\bmu,u^{-1}\bOmega)
	\Phi(u^{1/2} \blambda^\top(\mathbf{y}-\bmu))}
d H(u|\bnu),
\end{equation*}
where $\bOmega$ and $\blambda$ are given in \eqref{eqn siglambda}.

Depending on the distribution of the scale factor $U$ we have a different member of the SMSN family. For example, if $P(U=1)=1$ we have the skew normal distribution;
$U \sim \textrm{Gamma}(\nu/2,\nu/2)$, with $\nu>0$, corresponds to the \emph{skew Student-$t$ distribution} -- here we denote by  $\textrm{Gamma}(a,b)$ the {gamma} distribution with mean $a/b$ and variance $a/b^2$, with $a,b>0$; $U \sim \textrm{Beta}(\nu,1)$, with pdf  $f(u|\nu) =\nu  u^{\nu-1}$, $0 <u <1$,   $\nu>0$, corresponds to the  \emph{skew slash distribution}; If $U$ is binary with $P(U= \tau)=\rho=1-P(U=1)$, where  $0 < \tau, \, \rho <1$ (and therefore  $\bnu=(\tau,\rho)^\top$), we have the \emph{skew contaminated normal distribution}. Obviously, there are other distributions in the SMSN family, but for illustrative  purposes we restrict ourselves to these. The SMSN family, {first} defined by \cite{Branco_Dey}, includes the class of the  \emph{scale mixtures of normal (SMN) distributions}, defined by \cite{andrews_mallows_74} 
where {normality is assumed} for $\X$ in \eqref{eqn def SMSN} (and so  $\bDelta=\mathbf{0}$). In this case, we use the notations   $ \Y \sim  \textrm{SMN}_q(\bmu,\bSigma,\bnu)$ and $\textrm{SMN}_q(\cdot|\bmu,\bSigma,\bnu)$ for the respective pdf. 
Obviously,  this class contains the  normal, Student-$t$, slash and contaminated normal distributions.  

The skew Student-$t$ pdf is given by:
\begin{equation*}
\textrm{ST}_{q}(\mathbf{y}|\bmu,\bSigma,\bDelta,\nu)=
2\textrm{t}_q(\mathbf{y}|\bmu,\bOmega,\nu)\textrm{T}\left[\left(\frac{\nu+p}{\nu+(\y-\bmu)^\top\bOmega^{-1}(\y-\bmu)}\right)^{1/2}\blambda^\top 
(\mathbf{y}-\bmu)|\nu+p\right],
\end{equation*}
where $\textrm{t}_q(\cdot|\bmu,\bOmega,\nu)$ and $\textrm{T}(\cdot|\nu+p)$ denote,
respectively, the pdf of the $q$-variate Student-$t$ distribution with location  vector $\bmu$, scale  matrix $\bOmega$ and $\nu$ degrees of freedom, and the distribution function of the standard univariate Student-$t$ distribution with $\nu+p$ degrees of freedom, and $\bOmega$ and $\blambda$ are given in \eqref{eqn siglambda} -- for a proof, see \cite{Branco_Dey}. 

The skew slash distribution has pdf
\begin{equation*}\label{eqn densSSL}
\textrm{SSL}_{q}(\mathbf{y}|\bmu,\bSigma,\bDelta,\nu)=2\nu\int^1_0u^{\nu-1}\textrm{N}_p(\mathbf{y}|\bmu,u^{-1}\bOmega)\Phi(u^{1/2}
\blambda^\top(\mathbf{y}-\bmu))du, 
\end{equation*}
which  can be evaluated using the \texttt{R} function \verb"integrate" \citep{rmanual}, for example. 

The skew-contaminated normal distribution has pdf
\begin{align*}
& \textrm{SCN}_q(\mathbf{y}|\bmu,\bSigma,\bDelta,\bnu) =\\
& \quad{} 2\left\{\rho \textrm{N}_q(\y|\bmu,\tau^{-1}\bOmega) \Phi(\tau^{1/2}\blambda^\top (\y-\bmu)  )  + (1-\rho ) \textrm{N}_q(\y|\bmu,\bOmega) \Phi(\blambda^\top  (\y-\bmu)   \right\}, 
\end{align*}
which comes directly from the definition.

From Definitions \ref{def CFUSN} and \ref{def SMSN}, we have  that affine transforms of a SMSN distribution are  still SMSN. That is, if   $\C$ is an $ m \times q$ matrix with rank $m$, $\de$ is an $m$-dimensional vector  and $\Y \sim \textrm{SMSN}_q(\bmu,\bSigma,\bDelta,\bnu)$, then   $\C\Y+\de \sim \textrm{ SMSN}_m(\C\bmu  +\de,\C \bSigma \C^\top,\C \bDelta,\bnu)$.

A \emph{finite mixture of SMSN distributions with $G$ components} is defined by its pdf as  
\begin{equation} \label{eqn FM-SMSN}
g(\y|\bTheta)= \sum_{j=1}^{G}p_j \textrm{SMSN}_q(\y|\btheta_j),   
\end{equation}
where  $p_j \geq 0$ are   such that $\sum_j^G p_j=1$,   $\btheta_{j}= (\bmu_{j},\bSigma_{j},\bDelta_{j},\bnu_{j})$ and $\bTheta=(\btheta_{1},\ldots,\btheta_{G},p_{1},\ldots,p_{G})$. The pdf  $\textrm{SMSN}(\cdot|\btheta_{j})$ is named the $jth$ \emph{mixture component} and $p_j$ is the {corresponding} \emph{weight}. Hereafter, we call  \eqref{eqn FM-SMSN} as the  \emph{ FMSMSN model}. A hierarchical representation of this model is given by  $\Y | S=j \sim \textrm{SMSN}_q(\y|\btheta_j)$, where $S$ is a discrete latent variable with probability function $P(S=j)=p_j$, $j=1,\ldots,G$. It is interpreted as a classification variable: given that $S=j$ then we know that the underlying subject  came from a population with distribution $\textrm{SMSN}(\cdot|\btheta_{j})$. Then, using Definitions \ref{def CFUSN} and \ref{def SMSN},  we have the following hierarchical representation for $\Y$ distributed as  FMSMSN:    
\begin{align}
\Y|S=j, U=u, T=t  & \sim   \textrm{N}_{q}(\bmu_{j} + \bDelta_{j} t, u^{-1}\bSigma_{j});  \label{eqn RE FMSMSN  Y}\\
\quad T|U=u &\sim \textrm{TN}(0,u^{-1},(0,\infty));\label{eqn RE FMSMSN  T} \\
U &\sim  H(\cdot|\bnu_j) \label{eqn RE FMSMSN  U}\\
P(S=j)&=p_j, \quad   j=1,\ldots,G, \label{eqn variavel classificadora}
\end{align}
with $\textrm{TN}(\mu,\sigma^2,A)$ denoting a truncated normal distribution, which is the distribution of $W|W \in A$, where  $W \sim \textrm{N}(\mu,\sigma^2)$.  
% where $\mu$ and $\sigma^2$ are the mean and variance before truncation.  
This representation is useful to obtain {a} MCMC-type algorithm to perform posterior inference for the proposed model that will be presented next, and also   to generate artificial samples from a FMSMSN distribution. For more details about FMSMSN distributions, see \citet{LachosCabralZellerBook}. 

\section{The SMSN Mixture Measurement Error Model} \label{sec LMMSNMIX}
The ME model can be put in a more general setting, by  considering  a multivariate  unobserved response $\y=(y_1,\ldots,y_r)^\top$. Thus, we intend to model  the relationship between  $\y$ and $x$ by assuming that  
$$
\y = \balpha + \bbeta x,   
$$
where  $\balpha$ and $\bbeta$ are $r$-dimensional vectors of  unknown regression   parameters. Let  $p=r+1$ and  suppose  that $\y$ and $x$ are observed with error. What we actually observe is the $p$-dimensional random vector $\Z=(X,\Y^{\top})^{\top}$,  such that $X = x + \zeta$ and   $\Y= \y + \mathbf{e}$.
 Thus, 
\begin{equation} 
\Z  =\aee + \be x + \bepsilon,  \label{eqn affine transformations of Zi} 
\end{equation}
where  $\aee=(0,\balpha^{\top})^{\top}$, $\be=(1,\bbeta^{\top})^{\top}$,  $\bepsilon=(\zeta,\mathbf{e}^\top)^{\top}$,  $\zeta$ and $\mathbf{e}$ are errors when observing $x$ and $\y$, respectively.
 Alternatively, defining $\R=(x,\bepsilon^\top)^{\top}$, the ME model can be written as:
\begin{equation} 
\Z = \aee+\B\R,  \label{eqn_affine2}
\end{equation}
where $\B=[\be \,\,\, \mathbf{I}_{p}]$ is a $p \times (p+1)$ partitioned matrix with first column equal to $\be$ and $\mathbf{I}_{p}$ is the $p \times p$ identity matrix. 
In general,
it is supposed that  $x$, $\zeta$ and  $\mathbf{e}$ are independent, with   $x \sim  \textrm{N}(\mu,\gamma^2)$, $\zeta \sim  \textrm{N}(0,\omega_0^2)$ and  $\mathbf{e} \sim \textrm{N}_r(\mathbf{0},\bOmega_{e})$, where  $\bOmega_e=\textrm{diag}\{\omega^{2}_{1},\ldots,\omega^{2}_{r}\}$.   
Thus, 
\begin{equation}\label{eqn normal model}
\R  \sim  \textrm{N}_{1+p}\left((\mu,\mathbf{0}^{\top}_p)^{\top}, \textrm{block diag}\{ \gamma^2, \bOmega\} \right), \,\, \mbox{with }
\bOmega  =\textrm{block diag}\{\omega_0^2, \bOmega_e \}. 
\end{equation}

 As observed by \cite{galea.2005} and   \cite{vidal.2010}, since the first component of $\aee$  is equal to zero and the first component of $\be$ is equal to one, the model is identifiable.

The extension of the ME model proposed by  \cite{lachos_etal:09b,lachos_etal:10b} considers that $ x \sim   \textrm{SMSN}(\mu,\gamma^2,  \Delta,\bnu)$ and 
$\bepsilon \sim  \textrm{SMN}_{p}(\mathbf{0},\bOmega,\bnu)
$ are uncorrelated, where $\bOmega$ is given in \eqref{eqn normal model}.  
We propose to extend    their model,   by supposing that  $x$ has a FMSMSN distribution, such that 
\begin{equation*}
x| S=j \sim \textrm{SMSN}(\mu_{j},\gamma^2_{j},\Delta_{j},\bnu),  \,\,\,  j=1,\ldots,G,
\end{equation*}
where, like in \eqref{eqn variavel classificadora},  $P(S=j)=p_j$.  Thus, the  pdf of $x$ is the mixture  $\sum_{j=1}^{G} p_{j} \textrm{SMSN}(\cdot|\mu_{j},\gamma^2_{j},\Delta_{j},\nu). % \label{eqn marginal x eps}
$
We can write the  assumptions above as: 
\begin{equation*} %\label{eqn dist latent errors mix}
\R|S=j \sim \textrm{SMSN}_{1+p}\left[(\mu_{j},\mathbf{0}^{\top}_{p})^{\top}, \textrm{block diag}\{ \gamma^{2}_{j}, \bOmega\},(\Delta_{j},\mathbf{0}^{\top}_{p})^{\top},\bnu   \right],  
\end{equation*}
which implies that the marginal distribution of $\R$ is also  FMSMSN.  Note that we are supposing that  the scale factor parameter $\bnu$ is the same for all components of the mixture. This assumption is not so restrictive; see, for example, \cite{Cabral_Lachos_Madruga2012}, where the linear mixed model, which has a  similar structure,  is investigated.  We call this model the \emph{SMSN finite mixture measurement error model}, which will be denoted by FMSMSN-ME or FMSN-ME, FMST-ME, etc. if we use the specific distributions of the family. From \eqref{eqn RE FMSMSN  Y}, we have that  
\begin{align}
\R |(S=j, U=u, T=t) & \sim
\textrm{N}_{1+p}\left(( \mu_{j} + \Delta_{j} t,\mathbf{0}^{\top}_{p})^{\top},u^{-1} \textrm{block diag}\{ \gamma^2_{j}, \bOmega\} \right),\label{eqn joint dist latent errors} 
\end{align}
where the distributions of $(T,U)$ and $S$ are  given by \eqref{eqn RE FMSMSN  T}-\eqref{eqn RE FMSMSN  U} and \eqref{eqn variavel classificadora}, respectively. As  $\bepsilon|U=u \sim \textrm{N}_{p}(\mathbf{0},u^{-1}\bOmega)$, the distribution of 
the vector of observations  $\Z=(X,\Y^{\top})^{\top}$ has  the following stochastic representation, see Equation \eqref{eqn affine transformations of Zi}: 
\begin{align}
\Z|(x, U=u) &\sim  \textrm{N}_{p}(\aee+\be x,u^{-1}\bOmega); \nonumber  \\
x |(S=j, U=u, T=t) &\sim  \textrm{N}(\mu_{j} + \Delta_{j} t,u^{-1} \gamma^2_{j}), 
\nonumber 
%\label{eqn first stoch rep2}
\\
T|U=u &\sim \textrm{TN}(0,u^{-1},(0,\infty));\nonumber \\
U &\sim  H(\cdot|\bnu);\nonumber \\
P(S=j)&=p_{j},\,\,\,\,  j=1,\ldots,G. \label{eqn first stoch rep1}
\end{align}

An alternative representation can be obtained by integrating out the latent variable $x$. From Equations \eqref{eqn_affine2} and \eqref{eqn joint dist latent errors}, the first two equations of representation \eqref{eqn first stoch rep1} can be replaced with

\begin{equation} \label{eqn sec rep}
\Z|(S=j, U=u, T=t) \sim \textrm{N}_{p}\left(\aee+\mu_{j}\be+ \Delta_j \be t,
u^{-1}(\gamma_{j}^{2} \be \be^{\top} + \bOmega)\right).  
\end{equation}
These representations are useful to obtain {a} MCMC-type algorithm to perform posterior inference and also to simulate samples from the FMSMSN-ME model. 

Let  $\bTheta=(\balpha^{\top},\bbeta^{\top},\bmu^{\top},\bDelta^{\top},\bgamma^{\top},\bomega^{\top},\p^{\top},\bnu^{\top})^{\top}$ be the vector of parameters to be estimated, where $\bmu=(\mu_{1},\ldots,\mu_G)^{\top}$, $\bDelta=(\Delta_{1},\ldots,\Delta_{G})^{\top}$, $\bgamma=(\gamma^{2}_{1},\ldots,\gamma^{2}_{G})^{\top}$,  $\bomega=(\omega_{0}^{2},\omega^{2}_{1},\ldots,\omega^{2}_{r})^{\top}$ and $\p=(p_{1},\ldots,p_{G})^{\top}$. Denoting the conditional pdf of $\Z|\bTheta$ by  $\pi(\z|\bTheta)$,   Equations   \eqref{eqn RE FMSMSN  Y}-\eqref{eqn variavel classificadora} and \eqref{eqn sec rep} imply that  
\begin{equation} \label{eqn dens marginal FMSNMEM}
\pi(\z|\bTheta)=\sum_{j=1}^{G}p_{j} \textrm{SMSN}_{p}(\z|\bxi_{j},\bSigma_{j},\bLambda_{j},\nu),   
\end{equation}
where
\begin{align} 
\bxi_{j} & =  \aee+\mu_{j}\be, \,\,\,\, \bLambda_{j}= \Delta_{j} \be,  \,\,\, \mbox{and} \nonumber \\%\label{eqn xiLambdaSigma1}
\bSigma_j & =  \gamma_{j}^{2} \be \be^{\top} + \bOmega = \left( \begin{array}{cc} \gamma_{j}^2+\omega_0^2 & \gamma_{j}^2 \bbeta^{\top}\\ \gamma_{j}^2 \bbeta & \gamma_{j}^2 \bbeta \bbeta^{\top} + \bOmega_{e} \end{array} \right).  \label{eqn xiLambdaSigma2}
\end{align}
 
\section{Posterior inference} \label{sec mcmc estimation}
Let $\z_1,\ldots,\z_n$ be  an observed random  sample from the FMSMSN-ME model. The likelihood function is given by $\prod_{i=1}^n \pi(\z_i|\bTheta)$, where $\pi(\cdot|\bTheta)$ is  given in  Equation \eqref{eqn dens marginal FMSNMEM}. The prior specification for each of  the parameters  $\balpha$, $\bbeta$, $\bmu$ and $\bDelta$ is multivariate normal. Regarding the dispersion parameters in $\bgamma$, we adopt the hierarchical prior defined as  
$$
\gamma_{j}^{-2}|f \sim \textrm{Gamma}(e,f) \quad  j=1,\ldots,G; \quad f \sim \textrm{Gamma}(g,h).
$$
This hierarchical prior setup follows\,\cite{richardson_green_97}, where the univariate normal mixture case is investigated. Also, we fix $\omega_i^{-2} \sim \textrm{Gamma}(l,m)$,  $i=0,1,\ldots,r$. For the  vector of  weights, we apply the usual assumption 
$\p \sim  \textrm{Dir}(\kappa_{1},\ldots,\kappa_{G}),$ that is, a Dirichlet distribution with known positive hyperparameters. In all applications presented in this text, we have  chosen {hyperparameter} values of the prior distributions that express little prior knowledge. Thus, the prior covariance matrices of $\balpha$, $\bbeta$, $\bmu$ and $\bDelta$ are assumed to be diagonal with large variances, the hyperparameters  $g$, $h$, $l$ and $m$ are small  and positive  (in general we  fix the hyperparameter values of the gamma priors   equal to 0.01) and $\kappa_{1}=\cdots=\kappa_{G}=1$.

Each specific model in the SMSN  class has a scale factor  parameter $\bnu$ with specific interpretation,  deserving  a different treatment  for prior choice. For instance, there are several suggestions for {estimating} the unknown degrees of freedom of the  Student-$t$ model; see the discussions in\,\cite{Fonseca.Ferreira.Migon} and \cite{Garay.Bolfarine.Lachos.Cabral.2015}. Here, we do not treat this issue {in depth}, but adopt prior choices that have been  useful for our purposes. For example, for the FMST-ME model we fix as prior for $\nu$ an exponential distribution with parameter $\lambda >0$ with a second  level of hierarchy given by $\lambda \sim \textrm{U}(\lambda_0,\lambda_1)$ (a uniform distribution  on the interval $(\lambda_0,\lambda_1)$), where $0<\lambda_0<\lambda_1$. In general, we adopt $\lambda_0=0.04$ and $\lambda_1=0.5$. 
See \citet[p.~171]{congdon2007bayesian} for more details. For the FMSSL-ME model, $\nu \sim \textrm{Gamma}(\phi_{sl},\psi_{sl})$, where $\phi_{sl}$ and $\psi_{sl}$ are small and positive. For the FMSCN-ME model, a simple prior setup can be considered as $\rho \sim \textrm{beta}(\rho_0,\rho_1)$ and $\tau \sim \textrm{beta}(\tau_0,\tau_1)$, where $\rho_0$, $\rho_1$,  $\tau_0$ and $\tau_1$ are positive. In general, we adopt a uniform distribution as a prior for these parameters, that is, $\rho_0=\rho_1=\tau_0=\tau_1=1$.  Assuming  prior independence, the posterior distribution is  given by
\begin{align} 
\pi(\bTheta|\z_1,\ldots,\z_n) & \propto \left( \prod_{i=1}^n \pi(\z_i|\bTheta) \right) \pi(\balpha) \pi(\bbeta) \pi(\bmu) \pi(\bDelta) \pi(\p)\pi(\bnu|\lambda)\pi(\lambda) \nonumber  \\
& \quad {} \times \left(\prod_{j=0}^{r} \omega_j^2 \right)  \left(\prod_{j=1}^{G} \pi(\gamma_j^2|f) \right)  \pi(f). \label{eqn posterior}
\end{align}
The hyperparameter $\lambda$ must be dropped in the cases of the FMSSL-ME and FMSCN-ME models.

Because of the nonstandard form of the posterior distribution \eqref{eqn posterior}, the computation of posterior {moments} estimates is a very hard task. Also, it is not  easy to generate samples from this posterior using traditional Monte Carlo methods. A reliable alternative is to develop {a} MCMC-type algorithm. Using existing Bayesian software like  \texttt{JAGS} \citep{jags.paper} or \texttt{Stan} \citep{stan.paper}, this can be easily implemented  through representations \eqref{eqn first stoch rep1} and \eqref{eqn sec rep} -- the last one  is our choice to {carry out} the computations. Alternatively, these representations are useful to develop a Gibbs-type algorithm to be implemented using existing  software like \texttt{R} \citep{rmanual}. 

\section{Model Selection} \label{sec model selection}

The  \emph{deviance information criterion} (DIC) \citep{spiegelhalter2002bmm} is a common Bayesian tool to compare a  given set of candidate models. For a given sample 
$\z=(\z_1^\top,\ldots,\z_n^\top)^\top$  let 
\begin{equation} \label{eqn def deviance}
D(\z,\bTheta)=-2 \sum_{i=1}^n \log \pi(\z_i|\bTheta)
\end{equation}
be the \emph{deviance} -- for the FMSMSN-ME model the expression for $\pi(\cdot|\bTheta)$ is given in Equation \eqref{eqn dens marginal FMSNMEM}. 
%Suppose that  $\z$ is the  observed sample.
 In this case,  in order to simplify the notation, we write $D(\bTheta)$.  Let 
$$
\overline{D(\bTheta)}= \textrm{E}[D(\bTheta)|\z]=-2 \sum_{i=1}^{n}  \textrm{E}[\log \pi(\z_i|\bTheta)|\z]  
$$
be  the posterior mean deviance. The measure 
\begin{equation} \label{eqn tau eff}
\tau_{D} = \overline{D(\bTheta)} - D(\tilde{\bTheta}),   
\end{equation}
where $\tilde{\bTheta}$ is  an estimator of $\bTheta$, is called \emph{the effective dimension}. The DIC is defined as 
$
\textrm{DIC} = D(\overline{\bTheta})+2 \tau_{D}.  
$
The posterior mean $\overline{\bTheta}=\textrm{E}[\bTheta|\z]$ is a usual choice for $\tilde{\bTheta}$. In this case  the DIC has the following expression 
\begin{equation} \label{DIC definition}
\textrm{DIC} = -4 \sum_{i=1}^{n}  \textrm{E}[\log \pi(\z_i|\bTheta)|\z]+2 \sum_{i=1}^{n} \log \pi(\z_i|\overline{\bTheta}). 
\end{equation}

The terms $D(\overline{\bTheta})$ and $2 \tau_{D}$ are interpreted as a measure of fit and a penalty for model complexity, respectively. It is well known that there are some issues with this  definition. For example, $\tau_D$ is not invariant to reparameterizations.  That is, different parameterizations can produce different values of $\tau_D$, and {hence} different values of DIC. Also, in the mixture model case the posterior mean $\overline{\bTheta}$ can be a poor choice for $\tilde{\bTheta}$, mainly because 
the finite mixture likelihood is invariant under permutations of the component labels (this property of the likelihood  is usually called \emph{label switching}). If the prior is also invariant with respect to the labels,  all posterior means will be equal, and the plug-in mixture $\pi(\z_i|\overline{\bTheta})$ will have only one component.
As a consequence, the estimator $D(\overline{\bTheta})$ of $D({\bTheta})$ is unreasonable,  and expression \eqref{DIC definition} is useless. For more details, see the discussion in \citet[p.~13]{Stephens_Thesis}.

In the context of finite mixture models, a more applicable definition of DIC can be found in \cite{celeux2006dic}, see also  \cite{Spiegelhalter.JRSS.2014}. Observe that, while the statistics  $\pi(\z_i|\overline{\bTheta})$ {is} affected by  label switching, the \emph{posterior predictive density evaluated at $\z_i$}, given by  $\textrm{E}[\pi(\z_i|\bTheta)|\z]$,  is not.  Thus, it is more reasonable to consider  the latter as an  estimator of $\pi(\z_i|\bTheta)$  in expression \eqref{eqn def deviance}. Then, instead of  $D(\overline{\bTheta})$,    we use $-2 \sum_{i=1}^{n} \log  \textrm{E}[\pi(\z_i|\bTheta)|\z]$ as an  estimator of $D(\bTheta)$  in expression \eqref{eqn tau eff},  resulting in the following  alternative definition of DIC:  
\begin{equation*} 
\textrm{DIC}= -4 \sum_{i=1}^{n}  \textrm{E}[\log \pi(\z_i|\bTheta)|\z]+2 \sum_{i=1}^{n} \log  \textrm{E}[\pi(\z_i|\bTheta)|\z]. 
\end{equation*} 
Also, defining DIC in this way   provides  invariance to reparameterization.

In general it is a hard task to obtain  closed form expressions  for the posterior mean  $\textrm{E}[\log \pi(\z_i|\bTheta)|\z]$ and for  the posterior predictive density $\textrm{E}[\pi(\z_i|\bTheta)|\z]$, but these  integrals  can be easily approximated using  posterior MCMC samples.  Let 
$\bTheta^{(l)}$ be the  MCMC sample generated at the  $l$th step  of the algorithm,   $l=1,\ldots,L$. Then,  we have the following    approximation for the DIC: 
\begin{equation*} \label{eqn alterDIC}
-  \frac{4}{L} \sum_{l=1}^{L}  \sum_{i=1}^{n} \log \pi(\z_i|\bTheta^{(l)}) +   2 \sum_{i=1}^{n} \log \left(\frac{1}{L}\sum_{l=1}^{L} \pi(\z_i|\bTheta^{(l)}) \right).
\end{equation*}

\section{Simulation Studies} \label{sec aplication}
We present three simulation studies in order to show the applicability of our proposed method.

\subsection{Simulation Study 1 - Parameter Recovery} \label{sec recovery}
The aim  of this study is to analyze  the performance of the proposed method  by studying some frequentist properties of the estimates. In order to do so, an experiment was carried out as follows.  Fixing  $r=2$ and $G=2$, we first generated 100 {datasets} of size  $n$    from  the FMST-ME model with the following parameter setup: $\balpha = (0.4, 0.1)^\top$, $\bbeta = (0.8, 0.9)^\top$, $\omega_0^2=0.2$, $  \omega_1^2=0.3$,  $\omega_2^2= 0.4$, $\mu_1 = 2$, $\mu_2 = 8$, $\Delta_1 =-2$, $\Delta_2 =2$, $\gamma_1^2=\gamma_2^2=0.1$, $p_1 = 0.7$ and $\nu=3$. 

For each {dataset} and for each parameter, we obtained an approximation of the posterior mean estimate through MCMC samples. For this purpose, we drew 25,000 MCMC posterior samples with a burn-in of 5,000 iterations and thinning of 30 iterations. We considered the sample sizes $n=50, 100, 500$. Then, the experiment was repeated for the FMSN-ME, FMSSL-ME and FMSCN-ME models with the same parameter setup, except for  the FMSCN-ME model, in which case  we fixed $\rho=0.7$ and $\tau=0.3$. The   average   and   standard deviation values (in parentheses) computed across  100 posterior mean  estimates  are presented in Table~\ref{key}, where PV is the parameter value used to generate the {dataset}.     
\begin{table}[htbp]
	\scriptsize
	\renewcommand{\arraystretch}{0.9}
	\renewcommand{\tabcolsep}{0.1cm}
%	\vspace{-0.1cm}
	\caption{ Simulation study 1:     average   and   standard deviation values (in parentheses)  computed across  100 posterior mean  estimates of the parameters in the FMSMSN-ME model.  }\label{key}
%	\vspace{-0.1cm}
\begin{center}
	\begin{tabular}{c rrrrcrrr} \hline 
		&&&&&&& \\
		&                &                \multicolumn{3}{c}{FMSN-ME}     &&   \multicolumn{3}{c}{FMSCN-ME	}     \\\cline{3-5}   \cline{7-9}    \\    					
		Parameters          &     PV      &     \multicolumn{1}{c}{$n=50$} &		 \multicolumn{1}{c}{$n=100$}	 &		 \multicolumn{1}{c}{$n=500$}	 &&  \multicolumn{1}{c}{$n=50$} &		 \multicolumn{1}{c}{$n=100$}	 &		 \multicolumn{1}{c}{$n=500$}  \\
		&&&&&&& \\
		\cline{3-5} \cline{7-9}  \\  
		$\Delta_1$	           &	$-$2.0	         &	$-$1.786 (0.680)    &	$-$1.996 (0.302)   	   &	$-$2.008 (0.119)   & &$-$1.933 (0.965)      &	$-$2.206 (0.553)   	 &	$-$2.153 (0.322)   \\   
		$\Delta_2$	           &	2.0	                &	     1.578 (1.093)    &	        1.877 (0.741)   	&	       2.041 (0.172)   &  &     1.141 (1.817)   	 &	     2.108 (0.603)   	&	    2.143 (0.359)     \\
		$\omega_0^2$ 	       & 	0.2	               &	     0.207 (0.100)   &	        0.213  (0.062)   	&	       0.205 (0.026)   &  &      0.196 (0.138)   	 &	     0.217 (0.108)   	&	    0.226 (0.068)  \\
		$\omega_1^2$	       &	0.3	               &	     0.309 (0.085)    &	        0.313  (0.057)   	&	       0.301 (0.026)   & &      0.337 (0.161)   	 &	     0.372 (0.167)   	&	    0.338 (0.089)     \\
		$\omega_2^3$	       &	0.4	               &	     0.433 (0.114)    &	        0.419 (0.072)   	&	       0.410 (0.036)   & &      0.496 (0.244)   	 &	     0.509 (0.227)   	&	    0.452 (0.121)   \\
		$\alpha_1$	           &	0.4	               &	     0.391 (0.113)    &	        0.402 (0.082)   	&	       0.399 (0.038)   & &      0.415 (0.173)   	 &	     0.397 (0.124)   	&	    0.401 (0.054)  \\
		$\alpha_2$	           &	0.1	               &	     0.102 (0.134)    &	        0.100 (0.093)   	&	       0.100 (0.044)   & &      0.123 (0.201)   	 &	     0.114 (0.124)   	&	    0.105 (0.066)   \\
		$\beta_1$	            &	0.8	               &	      0.802 (0.020)    &	     0.802 (0.018)   	&	        0.800 (0.007)   &  &     0.795 (0.026)   	 &	     0.799 (0.021)   	&	    0.799 (0.008)   \\
		$\beta_2$	            &	0.9	               &	      0.897 (0.026)    &	     0.902 (0.019)   	&	        0.900 (0.008)   & &     0.892 (0.033)   	 &	     0.900 (0.024)   	&	    0.899 (0.009)   \\
		$\gamma^2_1$	&	0.1	               &	      0.207 (0.207)    &	     0.122 (0.152)   	&	        0.100 (0.063)   & &     0.216 (0.228)   	 &	     0.132 (0.180)   	&	    0.084 (0.077)  \\
		$\gamma^2_2$	&	0.1	               &	      0.222  (0.222)   &	     0.141  (0.160)   	&	        0.100 (0.071)   & &    0.258 (0.271)   	 &	     0.155 (0.200)   	&	    0.097 (0.085)   \\
		$\mu_1$	                &	2.0	               &	      1.787  (0.550)   &	     1.992 (0.230)   	 &	        2.003 (0.090)   & &    1.795 (0.774)   	 &	     2.042 (0.349)   	&	    2.038 (0.131)   \\
		$\mu_2$	                &	8.0	               &	      8.394 (0.807)    &	      8.121 (0.537)      &	        7.987 (0.114)   &   &    8.939 (1.709)   	 &	     8.099 (0.469)   	&	    7.973 (0.183)   \\
		$p_1$	                   &	0.7	              &	          0.694 (0.063)    &	      0.701 (0.043)   	&	       0.700 (0.020)   & &    0.680 (0.055)   	 &	     0.692 (0.043)   	&	    0.697 (0.018)\\
		$p_2$	                   &	0.3	              &	          0.305 (0.063)    &	      0.298 (0.043)   	&	       0.299 (0.020)  &   &    0.319 (0.055)   	 &	     0.307 (0.043)   	&	    0.303 (0.018)   \\
		$\rho$	                  &	0.7	                 &	-	                                 &	-	           	                      &-& &    0.668 (0.130)   	 &	     0.647 (0.128)   	&	    0.654 (0.104)   \\
		$\tau$	                  &	0.3	                 &	-	                                 &	-	                                  &	-&  &    0.320 (0.113)  	&	    0.354 (0.133)      &	   0.326 (0.069)   \\
		$\nu$	                  &	3.0	                 &	-	                                 &	-	                                  &-& &-	                                &	-	                              &	-	 \\
		\hline 	&&&&&&& \\
		&                &                \multicolumn{3}{c}{FMSSL-ME}     &&   \multicolumn{3}{c}{FMST-ME	}     \\\cline{3-5}   \cline{7-9}        				\\	
		Parameters          &     PV       &     \multicolumn{1}{c}{$n=50$} &		 \multicolumn{1}{c}{$n=100$}	 &		 \multicolumn{1}{c}{$n=500$}	 &&  \multicolumn{1}{c}{$n=50$} &		 \multicolumn{1}{c}{$n=100$}	 &		 \multicolumn{1}{c}{$n=500$} 	\\
		&&&&&&& \\
		\cline{3-5} \cline{7-9}   \\ 
		$\Delta_1$	           &	$-$2.0	         &	 $-$1.929 (0.758)   &	$-$2.100 (0.357)   &	$-$2.066 (0.178)   &&   $-$2.058 (0.539)   	 &	$-$2.078 (0.339)   &	 $-$2.045 (0.147)      \\   
		$\Delta_2$	           &	2.0	                &	      1.461 (1.585)   &	        1.988 (0.824)   &	       2.029 (0.214)   & &   1.903 (1.311)   	&	     2.079 (0.568)   &	        2.026 (0.187)     \\
		$\Omega_1$ 	       & 	0.2	                &	      0.233 (0.106)   &	        0.228 (0.087)   &	       0.211 (0.032)   & &   0.236 (0.124)   	&	     0.220 (0.088) 	 &	        0.207 (0.031)       \\
		$\Omega_2$	       &	0.3	                &	      0.334 (0.125)   &	        0.343 (0.088)   &	       0.310 (0.034)   & &   0.342 (0.122)   	&	     0.325 (0.089)   &	        0.303 (0.030)     \\
		$\Omega_3$	       &	0.4	                &	      0.453 (0.167)   &	        0.451 (0.112)   &	       0.415 (0.047)   & &  0.476 (0.167)   	&	     0.429 (0.114) 	 &	        0.411 (0.044)      \\
		$\alpha_1$	           &	0.4	                &	      0.389 (0.141)   &	        0.394 (0.110)   &	       0.396 (0.042)   & &  0.397 (0.145)   	&	     0.401 (0.092) 	 &	        0.392 (0.039)     \\
		$\alpha_2$	           &	0.1	                &	      0.109 (0.163)   &	        0.118 (0.115)   &	       0.106 (0.051)   & & 0.093 (0.142)   	&	     0.099 (0.107)	 &	        0.088 (0.054)        \\
		$\beta_1$	            &	0.8	                 &	       0.802 (0.024)   &	     0.798 (0.017) 	 &	        0.801 (0.007)   & &   0.796 (0.024)   	&	     0.801 (0.018) 	 &	        0.800 (0.007)      \\
		$\beta_2$	            &	0.9	                 &	       0.899 (0.030)   &	     0.897 (0.019)	 &	        0.900 (0.009)   & &  0.899 (0.030)   	&	     0.900 (0.020) 	 &	        0.901 (0.010)       \\
		$\gamma^2_1$	&	0.1	                 &	       0.117 (0.188)   &	     0.102 (0.174)   &	        0.081 (0.073)   & &   0.227 (0.208)   	&	     0.161 (0.189) 	 &	        0.092 (0.062)      \\
		$\gamma^2_2$	&	0.1	                 &	       0.144 (0.225)   &	     0.119 (0.187) 	 &	        0.094  (0.103)  & &     0.272 (0.238)   	&	     0.179 (0.207)   &	        0.104 (0.068)   \\
		$\mu_1$	                &	2.0	                 &	       1.845 (0.667)   	&	     2.019 (0.305)	 &	        2.033 (0.115)   & &  1.939 (0.350)   	&	     2.018 (0.207) 	 &	        2.029 (0.085)       \\
		$\mu_2$	                &	8.0	                 &	       8.580 (1.253)   	&	     8.097 (0.651)	 &	        7.978 (0.145)   & &  8.249 (0.969)   	&	     8.041 (0.466)   &	        7.982 (0.105)    \\
		$p_1$	                   &	0.7	                 &	       0.688 (0.054)   	&	     0.687 (0.047)   &	        0.699 (0.021)   & &  0.671 (0.060)   	&	     0.701 (0.045)	 &	        0.700 (0.019)     \\
		$p_2$	                   &	0.3	                 &	       0.311 (0.054)  	&	     0.312 (0.047)	 &	        0.301 (0.021)   & &  0.328 (0.060)   	&	     0.298 (0.045)	 &	        0.299 (0.019)   \\
		$\nu$	                  &	3.0	                     &	       6.308 (3.630)   &	 6.053 (3.580)    &	 3.554 (1.336)        & &  3.855 (1.380)   	    &	     3.478 (0.992)   &	 3.111 (0.311)  \\
		\hline  
	\end{tabular}
\end{center}
\end{table}

\begin{figure}[htbp] 
	\centering
	\includegraphics[scale=0.5,angle=-90]{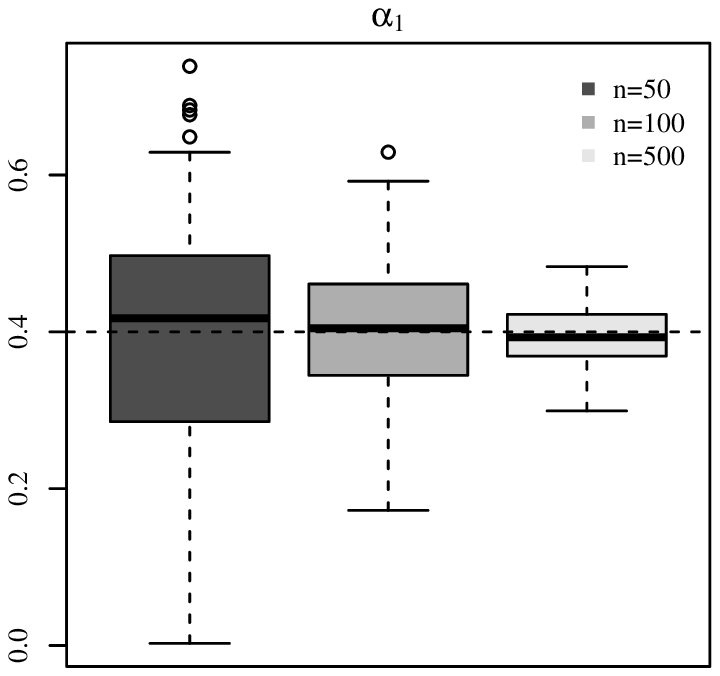}
	\includegraphics[scale=0.5,angle=-90]{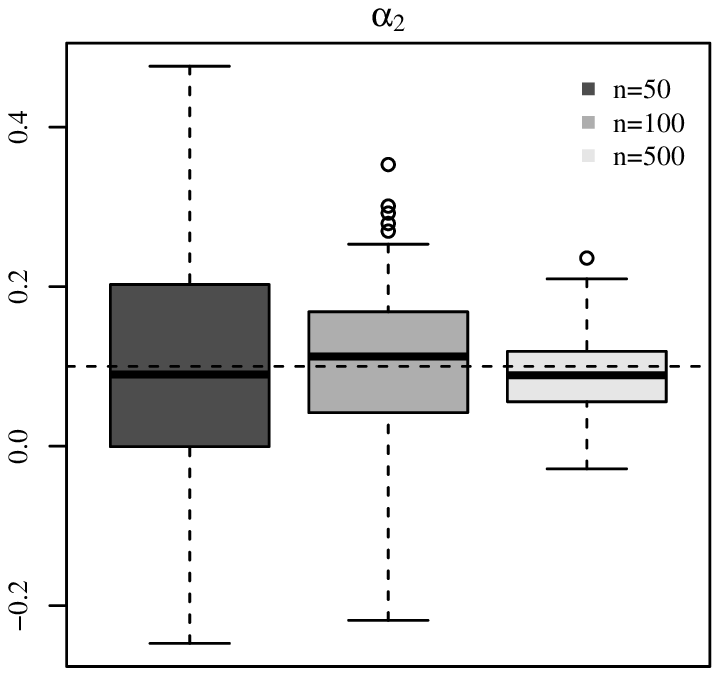}
	\includegraphics[scale=0.5,angle=-90]{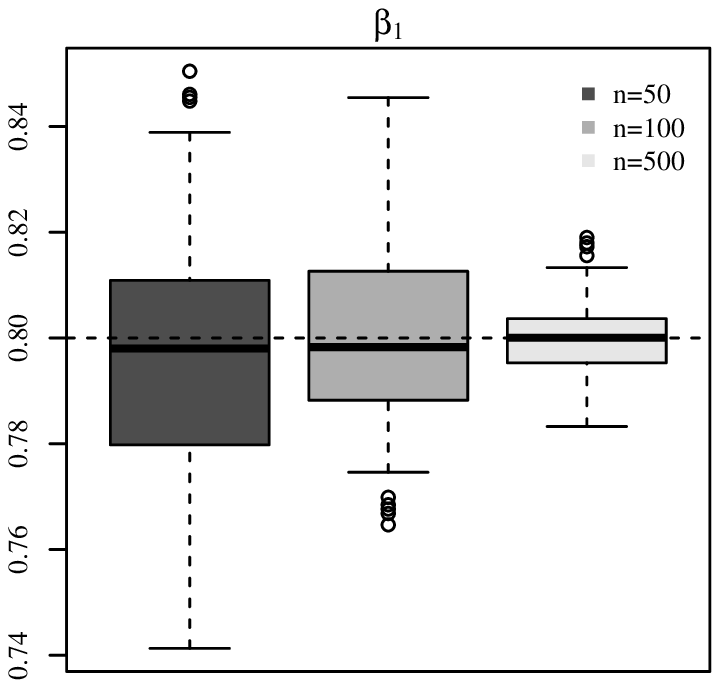}
	\includegraphics[scale=0.5,angle=-90]{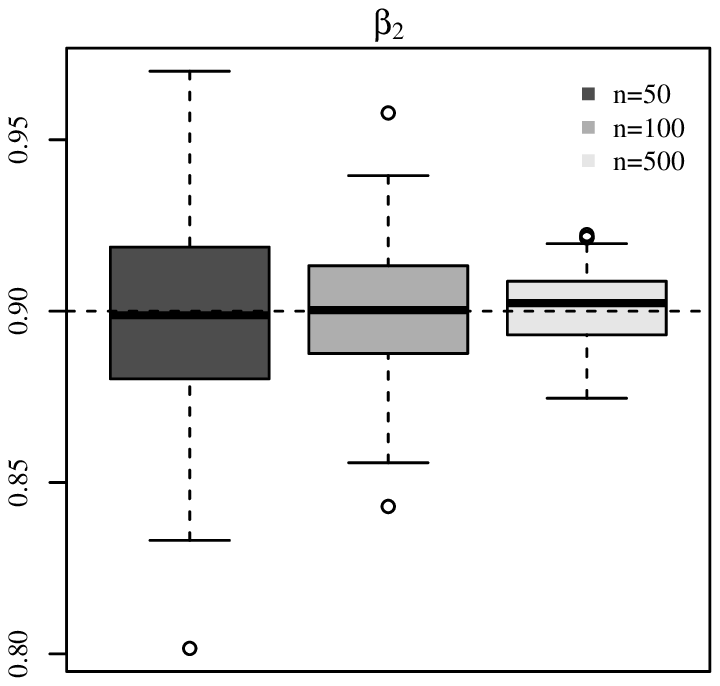}\\
	\includegraphics[scale=0.5,angle=-90]{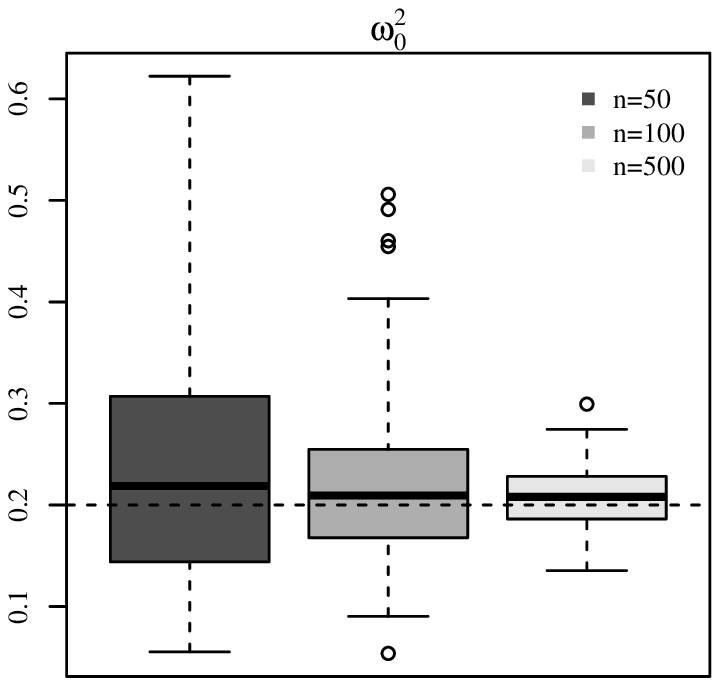}
	\includegraphics[scale=0.5,angle=-90]{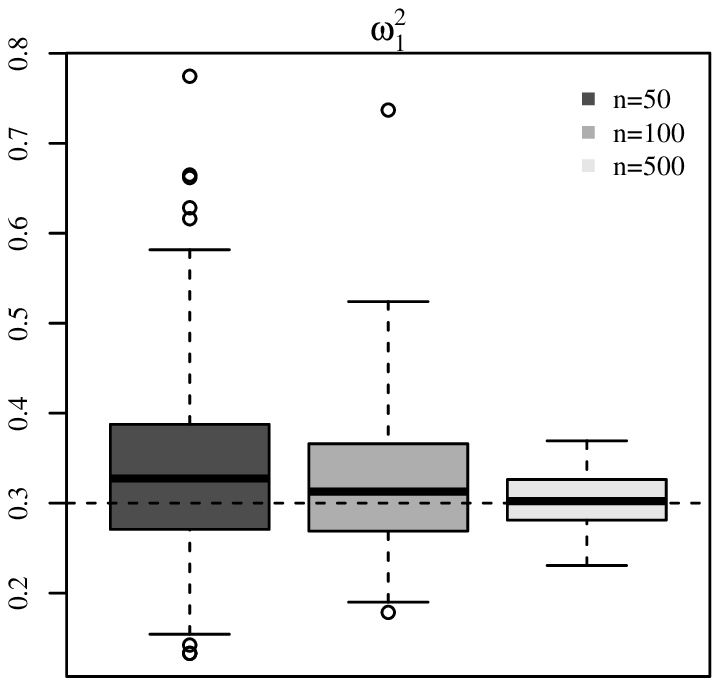}
	\includegraphics[scale=0.5,angle=-90]{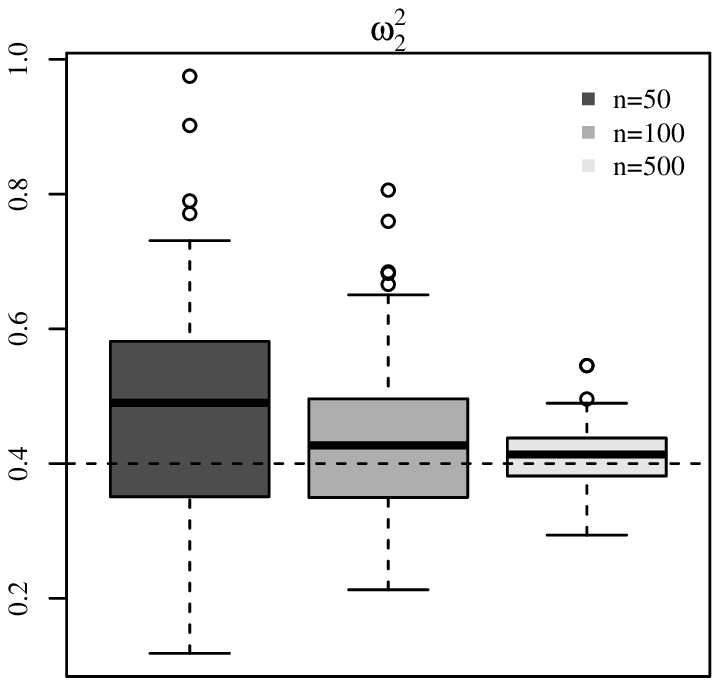}
	\includegraphics[scale=0.5,angle=-90]{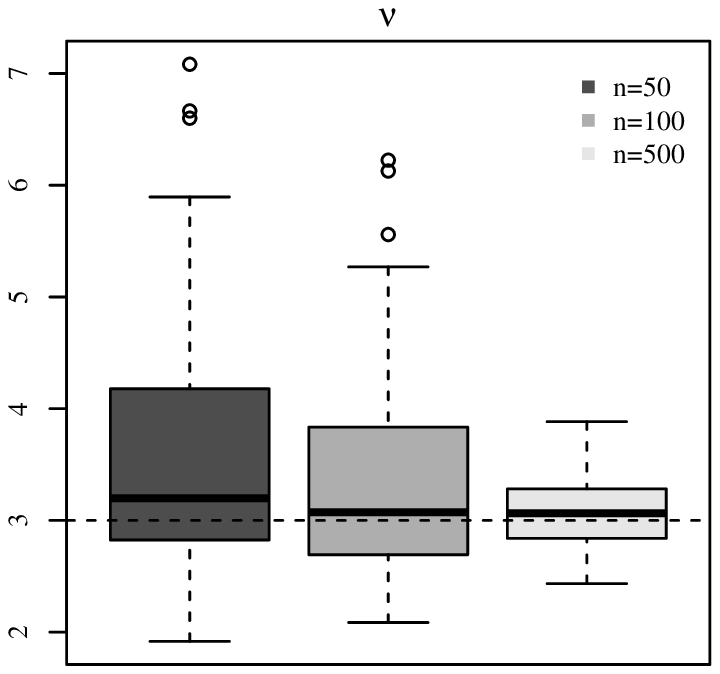}
	\vspace{-0.35cm}
	\caption{Simulation study 1. Boxplots of the estimates of the parameters in the FMST-ME model. The horizontal dotted lines   indicate the parameter value used to generate the {datasets}. } \label{box_st}
\end{figure}

The results are very satisfactory, even for the {relatively} small sample size $n=50$. This can be confirmed by {inspecting} some adjusted boxplots in Figure~\ref{box_st}. To save space  we only exhibit the boxplots for the FMST-ME case,  excluding the parameters of the mixture.

	\subsection{Simulation Study 2 - The Flexibility of the FMSMSN-ME model}
	
	{We now} study the performance of the proposed model when fitting data generated from a measurement error model with a latent covariate  having a distribution that is a finite mixture of normal inverse Gaussian distributions (NIG). This experiment is similar to that carried out by \cite{cabral_etal:14}; more details can be found in this reference. The main motivation is that the NIG distribution, contrary {to} SMSN class, is not a sub-family of the class of skew-elliptical distributions.

	The NIG  distribution is  a scale mixture of a normal distribution and an inverse Gaussian (IG) distribution. We say that a random variable $U$ has an IG distribution when its density is given by   
	$$ g(u)=
	\frac{\delta}{\sqrt{2 \pi}} u^{-3/2} \exp
	\left\{-\frac{1}{2}\left(\frac{\delta^2}{u}+ \gamma^2 u -2\delta
	\gamma \right) \right\}, \,\,\,\, u >0, $$ where $\gamma>0$ and
	$\delta>0$. In this case,  we  use the notation   $U \sim
	\textrm{IG}(\gamma,\delta)$.
	
	\begin{definition} \label{def NIG distribution}
		We say that the random vector $\X$ has a $p-$dimensional NIG
		distribution if it admits the representation
		\begin{eqnarray*}
			\X|U=u \sim \textrm{N}_p(\bmu+ u\bDelta \blambda ,u \bDelta ),
			\,\,\,\, U \sim \textrm{IG}(\gamma,\delta),
		\end{eqnarray*}
		where $\bmu$ and $\blambda$ are $p$-dimensional vectors of
		parameters, $\bDelta$ is a $p \times p$ positive definite matrix of parameters and 
		$\gamma$ and $\delta$ are positive parameters.
	\end{definition}
	
	We use the notation $\X \sim \textrm{NIG}(\bmu,\bDelta,\blambda,\gamma,\delta)$. It is assumed that $\textrm{det}(\bDelta)=1$; this restriction ensures identifiability. Observe that, when both $\gamma$ and $\delta$ tend to infinity, the limiting distribution is multivariate normal; see more details in  \citet{Barndorff-Nielsen1997}.
	
	{We now} define an alternative ME  model, by assuming that the marginal distribution of $x$ is a finite  mixture where the $jth$ component is NIG$(\mu_{j},1,\lambda_{j},\gamma_{j},\delta_{j})$, $j=1,\ldots,G$. We call this {the} FMNIG-ME model, defined  by 
	\begin{align*}
	\Z|(x, U=u) &\sim \textrm{N}_{p}\left(\aee+\be
	x,u\bOmega\right), \nonumber\\ x |(U=u,S=j) &\sim
	\textrm{ N}(\mu_{j} + u  \lambda_{j},u), \nonumber\\ U| S=j
	&\sim    \textrm{ IG}(\gamma_j,\delta_j), \quad j=1,\ldots,G, 
	%\label{eqn rep FMNIG-ME}
	\end{align*}
	where $ \textrm{det}(\bOmega)=1$. This definition is based on representation \eqref{eqn first stoch rep1}. The marginal distribution of $\Z$ is a mixture of NIG distributions, and each component $j=1,\ldots,G$ is distributed as  
	$
	\textrm{NIG}(\aee+ \be \mu_j,\be\be^{\top} + \bOmega,(\be\be^{\top} +
	\bOmega)^{-1} \be \lambda_j,\gamma_j,\delta_j). 
	$ 
	Samples of size $n=100$ and $n=500$ from the FMNIG-ME model with $p=3$ and $G=3$ were  generated with the following scenario: $p_1=0.4$, $p_2=0.3$, $\mu_1=-10$, $\mu_2=1$, $\mu_3=10$,  $\lambda_1=-2, \lambda_2=1$, $\lambda_3=-2$,  $\bOmega=\mathbf{I}_{3}$, $\balpha=(0.4,0.1)^{\top}$, $\bbeta=(0.8,0.9)^{\top}$,  $\delta_1=\delta_2=\delta_3=0.5$
	and $\gamma_1=\gamma_2=\gamma_3=1$. Then we fitted FMSMSN-ME models with three components to these data. We considered models based on symmetric distributions, namely FMN-ME (normal), FMT-ME (Student-$t$), FMSL-ME (slash), FMCN-ME (contaminated normal) and the {previously} cited skewed models, namely, FMSN-ME, FMST-ME, FMSSL-ME and FMSCN-ME models. In each case,  Table \ref{tab dic nig} presents the {DIC} values. {It can be seen} that the models that take into account skewness,  heavy tails and multi-modality at the same time outperform (values in boldface) the other ones.

	\begin{table}[htbp] 
		\centering
		\caption{DIC values   for the FMNIG-ME data.} \label{tab dic nig}
%		\vspace{-0.2cm}
		\begin{tabular}{lccc}
			\hline  
			& \multicolumn{3}{c}{Sample size}\\\cline{2-4}
			Model & {$n=100$} && {$n=500$}\\ \hline\\[-0.3cm]
			FMN-ME   &  970.2639 &&  5209.1670\\
			FMT-ME   &  945.2123 &&  4799.7490\\
			FMSL-ME  &  970.3589 &&  4923.7360\\
			FMCN-ME  &  978.9807 &&  4967.0020\\
			FMSN-ME  &  922.2715 &&  4951.6620\\
			FMST-ME  &   {\bf 874.2943} &&   {\bf 4546.9600}\\
			FMSSL-ME &  {\bf 885.3027}  &&  {\bf 4569.4180} \\
			FMSCN-ME &   {\bf 882.9718} &&  {\bf 4578.2080}\\	
			\hline
		\end{tabular}
	\end{table}

	\subsection{Simulation Study 3 - Identifiability}
	It is well known that identifiability is a sensitive issue for mixture of regression models. Few works deal with {this} problem, and the results  are restricted to some  specific models. Extensions of these results to models based on the SMSM class   are a challenge that, until now, has not been explored in {depth} in the literature. For a short discussion, see \cite{Zeller.Cabral.Lachos.2019}. Instead of a formal proof, we propose to study the identifiability of our proposed model using a simple method suggested by \cite{Lele.2010}, called \emph{data cloning}.   
	
	The data cloning algorithm allows us to approximate maximum likelihood estimates and the inverse of the Fisher information matrix using MCMC samples from a modified  posterior distribution of the vector of parameters in the model $\bTheta$. Let $\z=(\z_1^\top,\ldots,\z_n^\top)^\top$ be the observed sample and let $\z^{(K)}=(\z^\top,\ldots,\z^\top)^\top$ be the \emph{replicated data}, which {are obtained by replicating} the original data $K$ times.
	
	The vector $\z^{(K)}$ is seen as a result of a hypothetical experiment that replicates the original one $K$ times independently, yielding the same data $\z$ each time. Under suitable conditions, it is possible to show that, when $K$ is large, the posterior distribution of $\bTheta|\z^{(K)}$ is approximated by a normal distribution with mean equal to the maximum likelihood (ML) estimate $\widehat{\bTheta}$ and covariance matrix equal to $(1/K)  I^{-1}(\widehat{\bTheta})$, where $I(\bTheta)$ is the Fisher information matrix. Thus, the mean of MCMC samples drawn from the posterior distribution of $\bTheta|\z^{(K)}$ can be used to approximate   $\widehat{\bTheta}$,  and $K$ times the  covariance matrix   of these posterior samples can be used to approximate    the asymptotic covariance matrix  of $\widehat{\bTheta}$. 
	
	Also, \cite{Lele.2010} {showed that} if $g(\bTheta)$ is a function of the parameter vector $\bTheta$, and if the covariance matrix of the posterior distribution $g(\bTheta)|\z^{(K)}$ has its largest eigenvalue ${\lambda}_K$ converging to zero when $K$ increases, then $g(\bTheta)$ is estimable. This convergence to zero has the same
	rate as $1/K$. Let  $\widehat{\lambda}_K={\lambda}_K/{\lambda}_1$. 
	The authors recommend {detecting} this convergence feature by the analysis of a plot of $\widehat{\lambda}_K$ as a function of $K$, and {comparing} it with the expected value plot of $1/K$.
	
\begin{figure}[h] 
		\begin{center}
	\includegraphics[width=40pc,height=12pc]{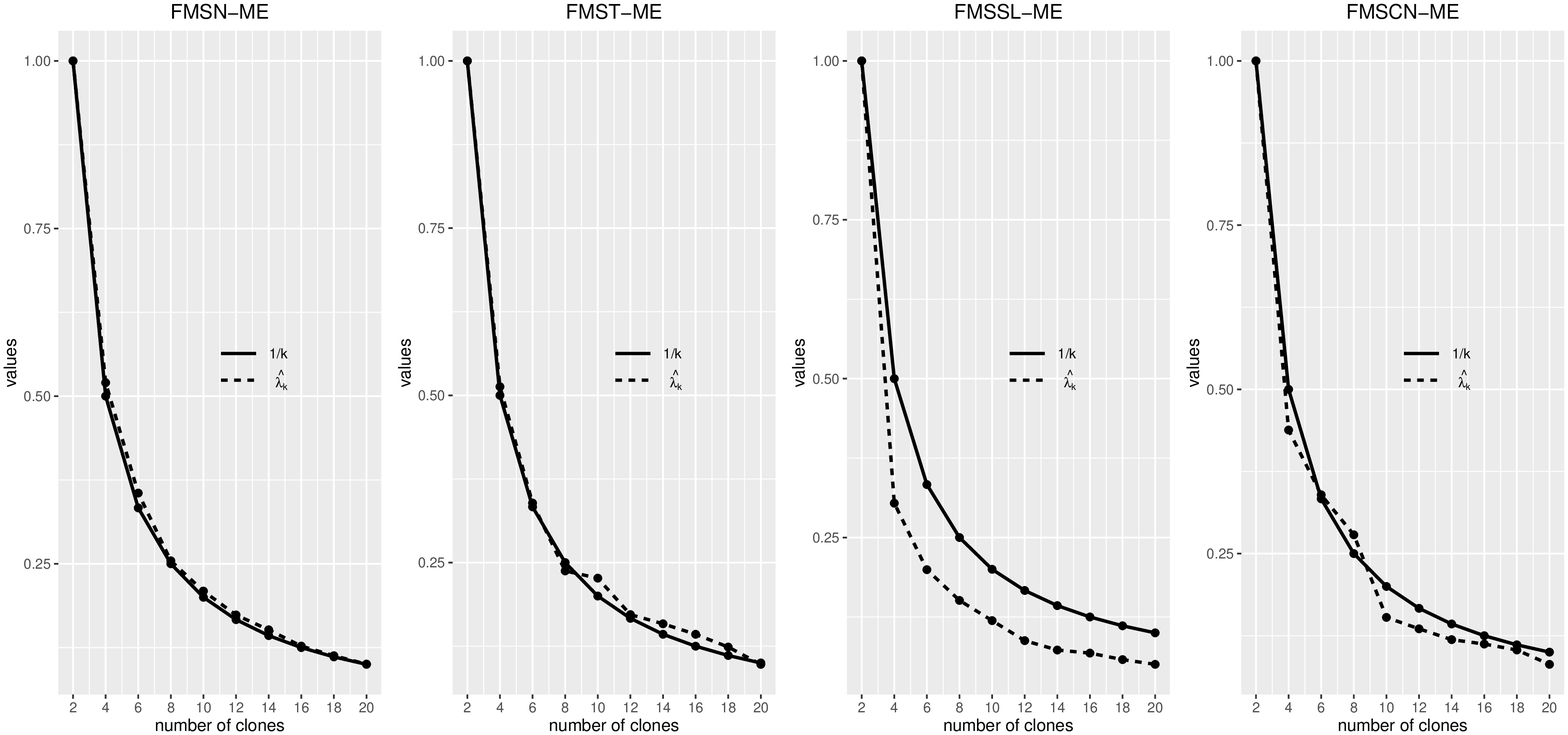}
	\vspace{-0.2cm}
	\caption{Simulation study 3. Identifiability checking using data cloning} 
	\label{data_cloning}
\end{center}
\end{figure}

	Figure~\ref{data_cloning} depicts these plots for the skewed models when $g(\bTheta)=\bTheta$, where $\z$ is an artificial sample generated using the same setup of Section~\ref{sec recovery}. {The procedure} was carried out using  the \texttt{R} package \texttt{dclone} \citep{Solymos.2010}. The plots suggest strong evidence of identifiability in all cases considered.

%	\begin{figure}[h] 
%		\includegraphics[scale=0.5,angle=-90]{dataClone-FMSN-ME}
%		\includegraphics[scale=0.5,angle=-90]{dataClone-FMST-ME}
%		\includegraphics[scale=0.5,angle=-90]{dataClone-FMSCN-ME}
%		\includegraphics[scale=0.5,angle=-90]{dataClone-FMSSL-ME}
%		\vspace{-0.2cm}
%		\caption{Simulation study 3. Identifiability checking using data cloning} 
%		\label{data_cloning}
%	\end{figure}

\subsection{Real {Dataset}} \label{sec_real_data}

\begin{figure} 
	\centering
%	\vspace{-0.25cm}
%	\includegraphics[scale=0.35]{hist24prot_ggplot}
%	\includegraphics[width=20pc,height=14pc]{hist24prot_ggplot}
%	\includegraphics[width=20pc,height=14pc]{histlupus_rpc}
%	\vspace{-0.25cm}
\includegraphics[scale=0.40]{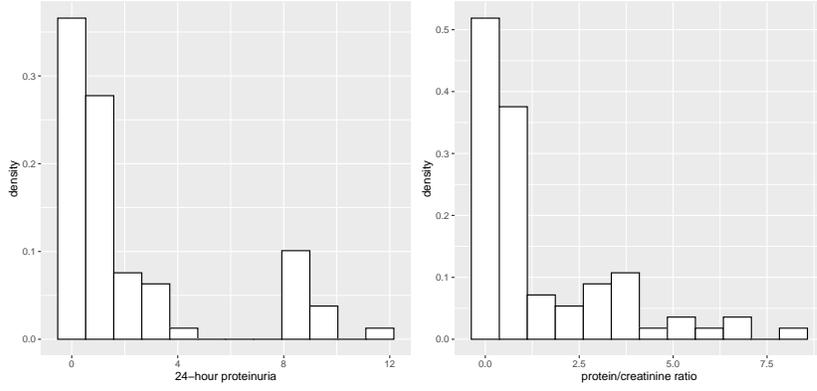}
	\caption{{Histograms for {protein}/creatinine ratio and 24-hour proteinuria   (both divided by 1000) for 75 SLE patients  }} \label{hist.data.lupos}
\end{figure}

We  illustrate our proposed methods with the SLE data described in Section~\ref{sec motivation}. In this case, $y$ and $x$ are the unobservable protein/creatinine ratio and   24-hour proteinuria, respectively. The respective measurements taken {from} 75 patients are denoted by $Y$ and $X$. The main goal is to study the relationship between these two tests. {Figure \ref{lupus_dispersion} shows that a} FMSMSN-ME model with two components can be a proper choice to model {these} data. This is confirmed by the visual inspection of Figure \ref{hist.data.lupos}. 
%\vspace{-0.25cm}
%\begin{figure}[htbp] 

We fitted FMSMSN-ME models with $G=1$ and $G=2$ to {these} data. {In this example we adopted a prior setup that was little different} from that defined in Section~\ref{sec mcmc estimation}, by fixing the hyperparameters of the prior distribution of $\rho$ in the FMCN-ME and FMSCN-ME models as $\rho_0=\rho_1=2$. 
%Figure \ref{data_MC} presents the traceplots of the MCMC draws for some parameters in the FMST-ME model, showing {evidence} of convergence of the algorithm. 
Table \ref{DIC.lupos} presents the DIC, the effective dimension and the  log-likelihood values (see Section \ref{sec model selection}) for the models. According to the DIC, the FMST-ME model with two components  is the best one. 

%\begin{figure}[htbp] 
%	\centering
%	\includegraphics[scale=0.45,angle=-90,page=1]{Estimate-ST}
%	\includegraphics[scale=0.45,angle=-90,page=2]{Estimate-ST}
%	\includegraphics[scale=0.45,angle=-90,page=3]{Estimate-ST}
%	\includegraphics[scale=0.45,angle=-90,page=4]{Estimate-ST}
%	\vspace{-0.2cm}
%	\caption{Traceplots for FMST-ME model, SLE data} 
%	\label{data_MC}
%\end{figure}

\begin{table}[htbp] 
%	\begin{table} 
	\centering
	\caption{Model selection for the SLE data} 
	\vspace{0.2cm}
	\begin{tabular}{lccc}\hline
		Model ($G$) & DIC & $\tau_D$ &  log-lik \\ \hline
		FMN-ME (1) & 693.7318 & 10.3267 & $-$341.7026\\
		FMST-ME (1) & 494.6313 & 6.2911 & $-$244.1701\\
		FMN-ME (2) &   556.1757  & 20.0197 & $-$268.078\\
		FMT--NE (2) &  544.4718  & 33.2336    & $-$255.6191   \\
		FMSL-ME (2) & 553.1619 & 19.7830 & $-$266.6895\\
		FMCN-ME (2) & 561.0051 & 26.4375 & $-$267.2838\\
		FMSN-ME (2) & 490.1661 & 12.0939 & $-$239.0361\\
		FMST-ME (2) & {\bf 483.7098} & 11.4611 & $-$236.1243\\
		FMSSL-ME (2) & 489.3703 & 11.2382 & $-$239.0661\\
		FMSCN-ME (2) & 502.7412 & 13.5445 & $-$244.5893\\
		\hline		
	\end{tabular}
	\label{DIC.lupos}
\end{table}

In order to study the fit of this model to the SLE data, we consider posterior predictive checking, by using the deviance $D(\cdot,\cdot)$ as a discrepancy measure between model and data -- see Equation \eqref{eqn def deviance} -- and computing the \emph{posterior predictive $p$-value (or Bayesian $p$-value)}, given by 

$$
p_{B}= P(D(\we,\bTheta) \geq D(\z,\bTheta)|\z)=\iint I_{A} \pi(\we|\bTheta) \pi(\bTheta|\z) d \we  d \bTheta, 
$$
where $\z=(\z_1^\top, \ldots,\z_n^\top)^\top$ is the observed sample, $\we=(\we_1^\top, \ldots,\we_n^\top)^\top$ is the replicated data that could have been observed and $A=\{(\we,\bTheta); \,\, D(\we,\bTheta) \geq D(\z,\bTheta)\}$. This $p$-value is the posterior probability that a future observation is more extreme (as measured by the deviance) than  the  data; see \citet[sec. 6.3]{Gelman.2014} for more details.   
Observe that $p_B$ is computed with respect to the joint  posterior distribution of $(\we,\bTheta)$ given $\z$.

\begin{figure}[htbp] 
	\centering
	\vspace{-0.2cm}
	\includegraphics[scale=0.35]{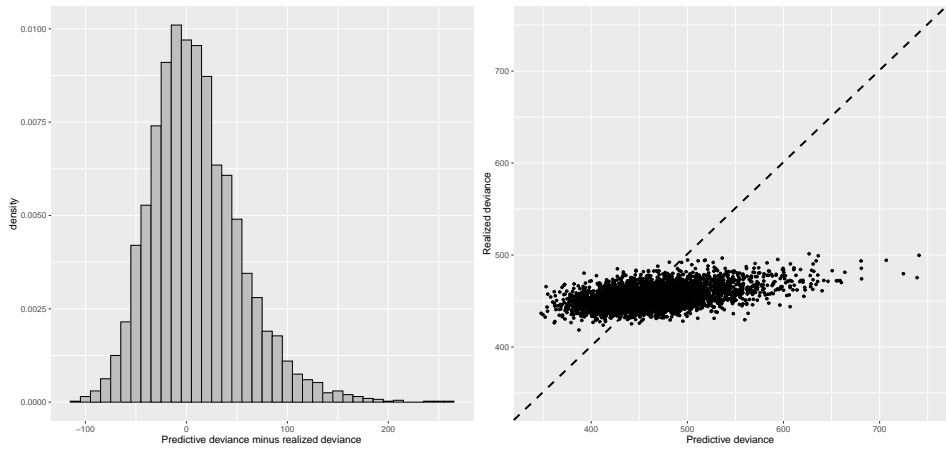}
	\vspace{-0.2cm}
	\caption{SLE data. Left side:  histogram of   $D(\we^{(l)},\bTheta^{(l)})$ (predictive deviance) minus $D(\z,\bTheta^{(l)})$ (realized deviance). Right side: scatterplot of predictive vs realized deviances. The p-value is computed as the proportion of points in the upper-left half of the scatterplot.
	}  \label{pvalor.lupos}
\end{figure}

\begin{figure}[htbp]
	\vspace{-0.2cm}
	\centering
	\includegraphics[scale=0.55]{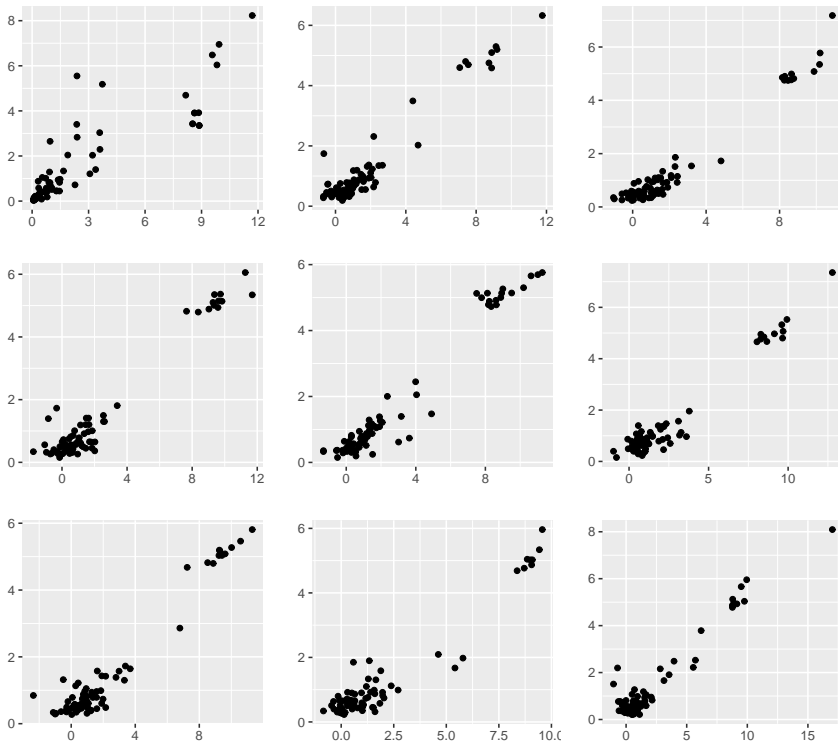}
	\vspace{-0.2cm}
	\caption{First row, first column: scatterplot of the SLE data. Other plots: scatterplots of replicated data} 
	\label{model check}
\end{figure}

It is possible to approximate $p_B$ using MCMC simulations. Let $\bTheta^{(l)}$ be the MCMC sample generated at the $l$th step of the algorithm, $l=1,\ldots,L$. Suppose that  $\we^{(l)}$ is drawn from $\pi(\cdot|\bTheta^{(l)})$,  which can be easily accomplished since this distribution is a mixture of SMSN distributions -- see Equation~\eqref{eqn dens marginal FMSNMEM}. Then the pairs $(\we^{(l)},\bTheta^{(l)})$, $l=1,\ldots,L$ are samples from the joint posterior distribution of $\we$ and $\bTheta$. Thus,  to approximate the Bayesian $p$-value, it is enough to observe the relative frequency  of  the event $A$  across the $L$ samples, that is,  the number of times $D(\we^{(l)},\bTheta^{(l)}) $ (the \emph{predictive deviance}) exceeds $D(\z,\bTheta^{(l)})$ (the realized deviance) out of the $L$ simulated draws. According
to \cite{Gelman.2014}, a model is suspect if a discrepancy is of practical importance and its $p$-value is  close to 0 or 1. In the case of the FMST-ME model with two components,  we obtained $p_{B} \approx 0.5455 $, {indicating no lack} of fit at all. Additionally, Figure \ref{pvalor.lupos} shows a histogram of the differences $D(\we^{(l)},\bTheta^{(l)})-D(\z,\bTheta^{(l)})$ and  a scatterplot of $D(\we^{(l)},\bTheta^{(l)})$ by $D(\z,\bTheta^{(l)})$. {Finally, Figure~\ref{model check}} presents a comparison between the actual data  and some replicated data, showing a close agreement between them. 
\vspace{-0.2cm}
%\begin{figure}[htbp] 
%	\centering
%	\vspace{-0.2cm}
%	\includegraphics[width=25pc,height=15pc]{model_check_3}
%	\vspace{-0.2cm}
%	\caption{SLE data. Left side:  histogram of   $D(\we^{(l)},\bTheta^{(l)})$ (predictive deviance) minus $D(\z,\bTheta^{(l)})$ (realized deviance). Right side: scatterplot of predictive vs realized deviances. The p-value is computed as the proportion of points in the upper-left half of the scatterplot.
%	}  \label{pvalor.lupos}
%\end{figure}

%\begin{figure}[htbp]
%	\vspace{-0.2cm}
%	\centering
%	\includegraphics[page=1,scale=0.5,angle=-90]{model_check}
%	\includegraphics[scale=0.5,angle=-90,page=2]{model_check}
%	\includegraphics[scale=0.5,angle=-90,page=3]{model_check}\\
%	\includegraphics[scale=0.5,angle=-90,page=4]{model_check}
%	\includegraphics[scale=0.5,angle=-90,page=5]{model_check}
%	\includegraphics[scale=0.5,angle=-90,page=6]{model_check}\\
%	\includegraphics[scale=0.5,angle=-90,page=7]{model_check}
%	\includegraphics[scale=0.5,angle=-90,page=8]{model_check}
%	\includegraphics[scale=0.5,angle=-90,page=9]{model_check}
%	\vspace{-0.2cm}
%	\caption{First row, first column: scatterplot of the SLE data. Other plots: scatterplots of replicated data} 
%	\label{model check}
%\end{figure}

\section{Conclusion}
In this article we proposed  an extension of the classical normal measurement error-in-variables model, flexible enough to accommodate at the same time skewness, heavy tails and multi-modality. Our approach is based on the joint modeling of the latent unobserved covariate and the random observational errors by a finite mixture of scale mixtures of skew-normal distributions. Stochastic representations of the model allow us to  develop  MCMC algorithms to perform  Bayesian estimation of the parameters in the proposed
model. Through the inspection of model selection criterion, simulated and real {datasets} were used to illustrate the advantages of our model over models based on symmetry. This method can be easily implemented using available software, making it useful for practitioners and researchers {in} several areas.  

\section*{Acknowledgments}

{
	The research was partially supported by CNPq and CAPES grants from the Brazilian federal government, and by FAPEAM grants from the government of the State of Amazonas, Brazil.
}

%\bibliographystyle{natbib}
%\bibliography{bibli}

\end{document}